\documentclass[twocolumn,showpacs]{revtex4}
\usepackage{amssymb}
\usepackage[dvips]{graphicx}

\input{tcilatex}
\begin{document}

\title{Key pairing interaction in layered doped ionic insulators}
\author{A. S. Alexandrov$^{1}$ and A. M. Bratkovsky$^{2}$}
\affiliation{Department of Physics, Loughborough University, Loughborough LE11 3TU,
United Kingdom\\
$^{2}$Hewlett-Packard Laboratories, 1501 Page Mill Road, Palo Alto,
California 94304}

\begin{abstract}
A controversial issue on whether the electron-phonon interaction (EPI) is
crucial for high-temperature superconductivity or it is weak and inessential
has remained one of the most challenging problems of contemporary condensed
matter physics. We employ a continuum RPA approximation for the dielectric
response function allowing for a selfconsistent semi-analytical evaluation
of the EPI strength, electron-electron attractions, and the carrier mass
renormalisation in layered high-temperature superconductors. We show that
the Fr\"{o}hlich EPI with high-frequency optical phonons in doped ionic
lattices is the key pairing interaction, which is beyond the
BCS-Migdal-Eliashberg approximation in underdoped superconductors, and it
remains a significant player in overdoped compounds.
\end{abstract}

\maketitle

For a long time, a basic question concerning the key pairing interaction in
cuprate and other high-temperature superconductors has remained open. Some
density functional (DFT) calculations \cite{cohen,heid} found small EPI
insufficient to explain high critical temperatures, $T_{c}$, while some
other first-principles studies found large EPI in cuprates \cite{bauer} and
in recently discovered iron-based compounds \cite{yndurain}. It is a common
place that DFT underestimates the role of the Coulomb correlations,
predicting an anisotropy of electron-response functions much smaller than
that experimentally observed in the layered high-$T_{c}$ superconductors.
The inclusion of a short-range repulsion (Hubbard $U$) via the LDA+U
algorithm \cite{zhang} and/or nonadiabatic effects \cite{bauer}
significantly enhances the EPI strength due to a poor screening of some
particular phonons. Also substantial isotope effects on the carrier mass and
a number of other independent observations (see e.g.\cite{asazhao} and
references therein) unambiguously show that lattice vibrations play a
significant although unconventional role in high-temperature
superconductors. Overall, it seems plausible that the true origin of
high-temperature superconductivity could be found in a proper combination of
strong electron-electron correlations with a significant EPI \cite{acmp}.

Here, we calculate the EPI strength with optical phonons, the phonon-induced
electron-electron attraction, and the carrier mass renormalisation in
layered superconductors at different doping using a continuum approximation
for the renormalised carrier energy spectrum and the RPA dielectric response
function. The Fr\"{o}hlich EPI with high-frequency optical phonons turns out
the key pairing interaction in underdoped highly polarizable ionic lattices
and remains significant player in overdoped compounds.

We start with a parent insulator as La$_{2}$CuO$_{4}$, where the magnitude
of the Fr\"{o}hlich EPI is unambiguously estimated using the static, $%
\epsilon _{s}$ and high-frequency, $\epsilon _{\infty }$ dielectric
constants \cite{ale96,alebra}. To assess its strength, one can apply an
expression for the polaron binding energy (polaronic level shift) $E_{p}$,
which depends only on the measured $\epsilon _{s}$ and $\epsilon _{\infty }$,%
\begin{equation}
E_{p}={\frac{e^{2}}{{2\epsilon _{0}\kappa }}}\int_{BZ}{\frac{d^{3}q}{{(2\pi
)^{3}q^{2}}}}.  \label{shift}
\end{equation}%
Here, the integration goes over the Brillouin zone (BZ), $\epsilon
_{0}\approx 8.85\times 10^{-12}$ F/m is the vacuum permittivity, and $\kappa
=\epsilon _{s}\epsilon _{\infty }/(\epsilon _{s}-\epsilon _{\infty })$. In
the parent insulator, the Fr\"{o}hlich interaction alone provides the
binding energy of two holes, $2E_{p}$, an order of magnitude larger than any
magnetic interaction ($E_{p}=0.647$ eV in La$_{2}$CuO$_{4}$ \cite{alebra}).
Actually, Eq.(\ref{shift}) underestimates the polaron binding energy, since
the deformation potential and/or molecular-type (e.g. Jahn-Teller \cite{mul}%
) EPIs are not included.

It has been argued earlier \cite{ale96} that the interaction with c-axis
polarized phonons in cuprates remains strong also at finite doping due to a
poor screening of high-frequency electric forces as confirmed in some
pump-probe \cite{boz,dragan} and photoemission \cite%
{shen,shentheory,meevasana} experiments. However, a quantitative analysis of
the doping dependent EPI has remained elusive because the dynamic dielectric
response function, $\epsilon (\omega ,\mathbf{q})$ has been unknown.

Recent observations of the quantum magnetic oscillations in some underdoped
\cite{und} and overdoped \cite{over} cuprate superconductors are opening up
a possibility for a quantitative assessment of EPI in these and related
doped ionic lattices with the quasi two-dimensional (2D) carrier energy
spectrum. The oscillations revealed cylindrical Fermi surfaces, enhanced
effective masses of carriers (ranging from $2m_{e}$ to $6m_{e}$) and the
astonishingly low Fermi energy, $E_F$, which appears to be well below 40 meV
in underdoped Y-Ba-Cu-O \cite{und} and less or about 400 meV in heavily
overdoped Tl2201 \cite{over}. Such low Fermi energies \cite{ref} make the
Migdal-Eliashberg (ME) adiabatic approach to EPI \cite{mig} inapplicable in
these compounds. Indeed, the ME non-crossing approximation breaks down at $%
\lambda \hbar \omega _{0}/E_{F}>1$ when the crossing diagrams become
important. The characteristic oxygen vibration energy is about $\hbar \omega
_{0}=80$ meV  in oxides \cite{pickett,timusk}, so that the ME theory cannot
be applied even for a weak EPI with the coupling constant $\lambda <0.5$. In
the strong coupling regime, $\lambda \gtrsim 0.5,$ the effective parameter $%
\lambda \hbar \omega _{0}/E_{F}$ becomes large irrespective of the adiabatic
ratio, $\hbar \omega _{0}/E_{F}$, because the Fermi energy shrinks
exponentially due to the polaron narrowing of the band \cite{ale83}. Since
carriers in cuprates are in the non-adiabatic (underdoped) or near-adiabatic
(overdoped) regimes, $E_{F}\lesssim \hbar \omega _{0}$, their energy
spectrum renormalised by EPI can be found with the familiar small-polaron
canonical transformation at \emph{any coupling }$\lambda $ \cite{alekor}.

The matrix element of the screened electron-phonon (Fr\"{o}hlich)
interaction is found as \cite{mahan}:
\begin{equation}
\gamma (\mathbf{q})={\frac{\gamma _{0}(q)}{{\epsilon (\omega _{0},\mathbf{q})%
}}},
\end{equation}%
where $\gamma _{0}(q)$ is the bare (unscreened) vertex in the parent
insulator, Fig.1a. In our selfconsistent approach $\epsilon (\omega ,\mathbf{q%
})$ is calculated in the loop (RPA) approximation \cite{ref0} with the exact
(polaronic) carrier propagators taking into account the phonon
\textquotedblleft dressing" of carries:
\begin{equation}
\epsilon (\omega ,\mathbf{q})=1+{\frac{2e^{2}}{{\epsilon _{0}\epsilon
_{\infty }q^{2}\Omega}}}\sum_{\mathbf{k}}{\frac{f_{\mathbf{k+q}/2}-f_{%
\mathbf{k-q}/2}}{{\hbar (\omega +i/\tau )-\epsilon _{\mathbf{k+q}%
/2}+\epsilon _{\mathbf{k-q}/2}}}}.  \label{loop}
\end{equation}%
Here, $\epsilon _{\mathbf{k}}$ is the polaron band dispersion, $\tau $ is
the relaxation time, $\Omega$ is the normalization volume, and $f_{\mathbf{k}%
}$ the Fermi-Dirac distribution function. The effect of collisions cannot be
always taken into account merely by replacing $\omega $ by $\omega +i/\tau $%
, in the collision-less Lindhard dielectric function, Eq.(\ref{loop}), in
particular at low frequencies or in the static limit \cite{kliewer,mermin},
where ladder-type vertex corrections are important \cite{ando}. However, in
our case the relevant frequency is the (renormalised) optical phonon
frequency, so that the vertex corrections are negligible as $1/\omega
_{0}\tau \ll 1$.
\begin{figure}[tbp]
\begin{center}
\includegraphics[angle=-00,width=0.53\textwidth]{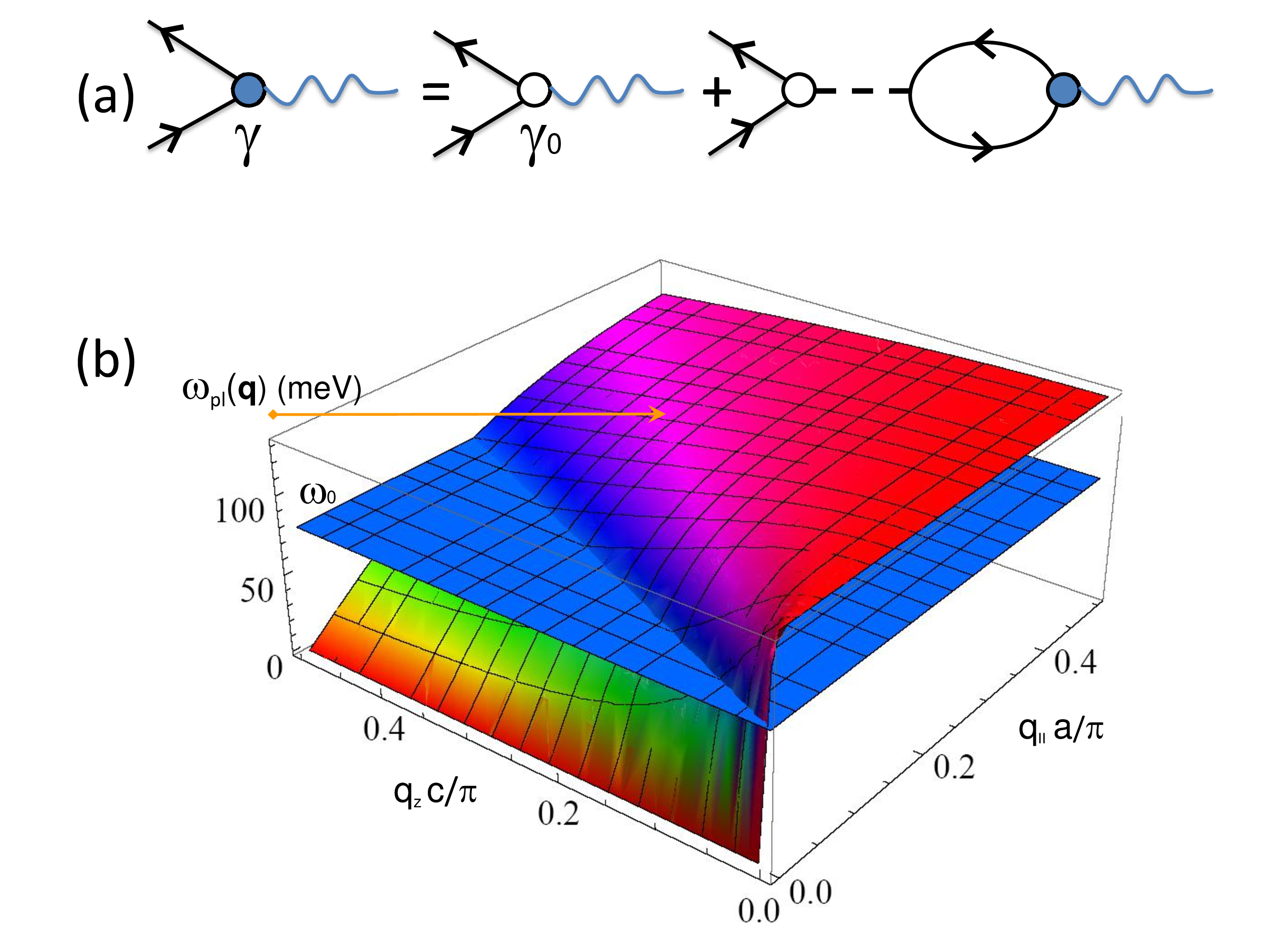} \vskip -0.5mm
\end{center}
\caption{(Color online) Diagrammatic representation of the screened EPI
vertex (a): solid lines correspond to polaron propagators, wavy lines are
the exact phonon propagator, and the dashed line is the Coulomb repulsion.
(b): long-wave dispersion of zeros of the dielectric function of quasi-2D
carriers with the 3D Coulomb repulsion. An optical phonon with the energy $80
$ meV is also shown.}
\label{brunch}
\end{figure}

Since the Fermi surfaces measured in the quantum oscillation experiments
\cite{und,over} are almost perfect cylinders, one can apply the continuum
(parabolic) approximation for the quasi-2D polaron energy spectrum, $%
\epsilon _{\mathbf{k}}=\hbar ^{2}(k_{x}^{2}+k_{y}^{2})/2m^{\ast }$, where
the polaron effective mass, $m^{\ast }$ has to be found self-consistently as
a function of EPI. Calculating the sum in Eq.(\ref{loop}) yields the
following dielectric response function extending the familiar pure 2D
collision-less result by Stern \cite{stern} to the 3D Coulomb interaction of
carriers with the quasi-2D energy spectrum and collisions,
\begin{equation}
\epsilon (\omega ,\mathbf{q})=1+{\frac{Ne^{2}}{{\epsilon _{0}\epsilon
_{\infty }q^{2}m^{\ast }v_{F}^{2}}}}[\chi _{1}(\omega ,q_{\parallel })+i\chi
_{2}(\omega ,q_{\parallel })].  \label{epsilon}
\end{equation}%
Here and below $q^{2}=q_{z}^{2}+q_{\parallel }^{2}$ is the square of the
phonon momentum, $N=k_{F}^{2}/2\pi c$ is the carrier density, $a$ and $c$
are in-plane and c-axis (chemical) unit cell constants, respectively, and $%
v_{F}=\hbar k_{F}/m^{\ast }$ is the Fermi velocity. The real and imaginary
parts of susceptibility are found as (see also \cite{connel}) $\chi
_{1}(\omega ,q_{\parallel })=2-[R(z,z^{2}-u,\beta )+R(z,z^{2}+u,\beta
)]/z^{2}$ and $\chi _{2}(\omega ,q_{\parallel })=[I(z,z^{2}-u,\beta
)-I(z,z^{2}+u,\beta )]/z^{2}$, where
\begin{equation}
{\frac{R(z,y,\beta )}{{\mathrm{sign}(y)}}}= \left[ {\frac{y^{2}-z^{2}-\beta
^{2}+\sqrt{(y^{2}-z^{2}-\beta ^{2})^{2}+4\beta ^{2}y^{2}}}{{2}}}\right]
^{1/2},  \nonumber
\end{equation}
\begin{equation}
I(z,y,\beta )=\left[{\frac{z^{2}+\beta ^{2}-y^{2}+\sqrt{(z^{2}+\beta
^{2}-y^{2})^{2}+4\beta ^{2}y^{2}}}{{2}}}\right] ^{1/2},
\end{equation}%
$z=q_{\parallel }/2k_{F}$, $u=\omega /(2k_{F}v_{F})$ and $\beta =1/(2k_{F}l)$
($l=v_{F}\tau $ is the mean-free path).

As shown in Fig.1b, a collective mode, $\omega _{pl}(\mathbf{q})$, defined
by $\epsilon (\omega _{pl},\mathbf{q})=0$ appears within the same frequency
range as the optical phonon mode $\omega _{0}$, leading to the
plasmon-phonon mixing (so-called plasphons \cite{ale92}). In the long wave
limit we find using Eq.(\ref{epsilon}), $\omega _{pl}(\mathbf{q})\approx
\omega _{p}q_{\parallel }/q$ for collision-less carriers ($\beta =0$), where
$\hbar \omega _{p}=(e^{2}E_{F}/4\pi \epsilon _{0}\epsilon _{\infty }c)^{1/2}$
is c.a. $132$ meV for $E_{F}=40$ meV . This mode is softer when it
propagates \emph{across} the planes than \emph{along} the planes, Fig.1b,
due to a low susceptibility of quasi-2D carriers to the electric field
applied across the planes \cite{ref00}.

\begin{figure}[tbp]
\begin{center}
\includegraphics[angle=-00,width=0.53\textwidth]{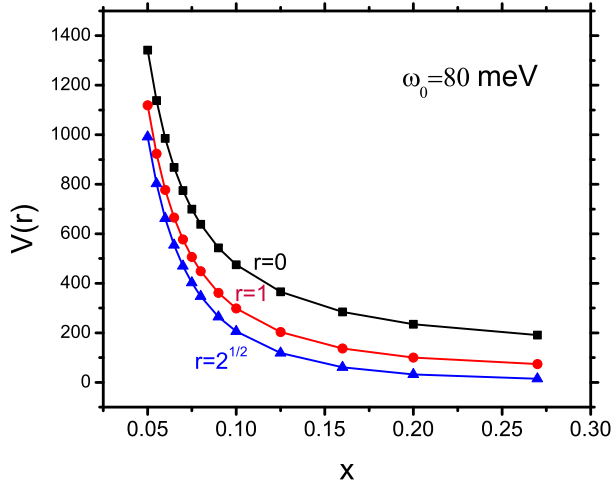} %
\includegraphics[angle=-00,width=0.53\textwidth]{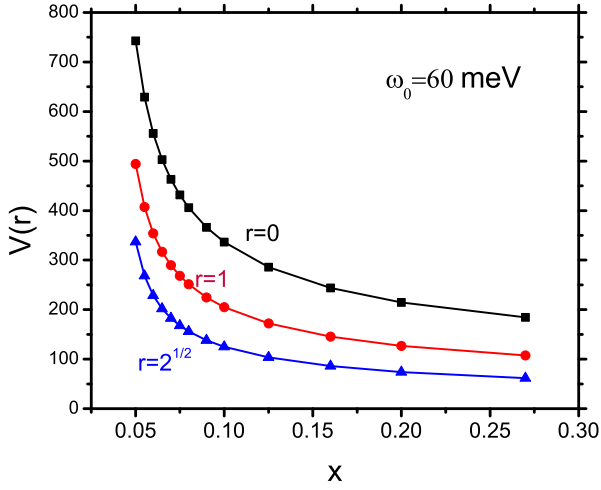}
\end{center}
\caption{(Color online) The on-site, $V(0)$ (upper curve), the
nearest-neighbor, $V(1)$ (middle curve), and the next-nearest-neighbor, $V(%
\protect\sqrt{2})$ (lower curve), attractions induced by the Fr\"{o}hlich
EPI for different doping $x$ and two characteristic phonon frequencies, $%
\hbar\protect\omega_0=80$ meV and $\hbar\protect\omega_0=60$ meV. }
\label{attractionfig}
\end{figure}
The polaron level shift, $E_{p}=V(0)/2$, the carrier attraction induced by
EPI, $-V(\mathbf{r})$, the in-plane polaron mass and the mass
renormalisation exponent, $g^{2}$, are found as \cite{aledev}
\begin{equation}
E_{p}={\frac{e^{2}}{{2\epsilon _{0}\kappa (2\pi )^{3}}}}\int_{BZ}{\frac{%
d^{3}q}{{q^{2}|\epsilon (\omega _{0},\mathbf{q})|^{2}}}},
\end{equation}%
\begin{equation}
V(\mathbf{r})={\frac{e^{2}}{{\epsilon _{0}\kappa (2\pi )^{3}}}}\int_{BZ}{%
\frac{d^{3}q\cos (\mathbf{r\cdot q})}{{q^{2}|\epsilon (\omega _{0},\mathbf{q}%
)|^{2}}}},
\end{equation}%
\begin{equation}
m^{\ast }=m\exp (g^{2}),  \label{mass}
\end{equation}%
and
\begin{equation}
g^{2}={\frac{E_{p}-V(\vec{j})/2}{{\hbar \omega _{0}}}},  \label{g}
\end{equation}%
respectively, where $\vec{j}$ connects nearest-neighbor sites. Holes in
cuprates reside on oxygen, so that the nearest-neighbor hopping distances is
$j=a/\sqrt{2}$. The BCS coupling constant with phonons is defined as $%
\lambda =E_{p}ma^{2}/\pi \hbar ^{2}$ in the case of 2D carriers with a
constant density of states ($ma^{2}/2\pi \hbar ^{2}$ per spin), where $m$ is
the bare band mass in a rigid lattice. Using $E_p \gtrsim 0.6 $eV and $m=2m_e
$ places cuprates in the strong-coupling regime, $\lambda \gtrsim 0.86$.
\begin{figure}[tbp]
\begin{center}
\includegraphics[angle=-00,width=0.53\textwidth]{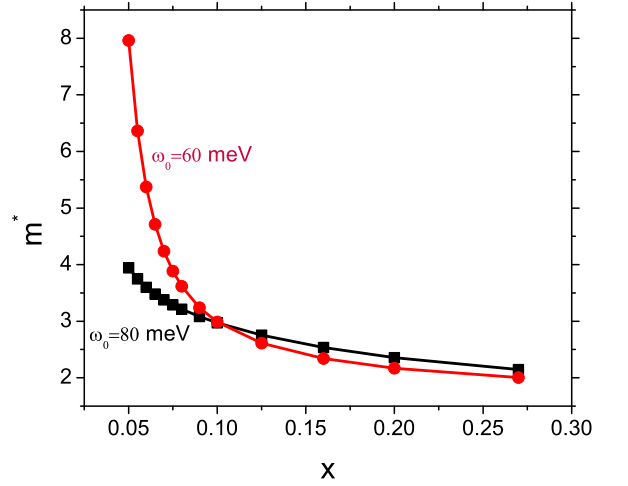}
\end{center}
\caption{(Color online) The carrier mass (in units of the band mass $m=m_e$)
as a function of doping for two characteristic phonon frequencies, $\hbar%
\protect\omega_0=80$ meV and $\hbar\protect\omega_0=60$ meV.}
\label{massfig}
\end{figure}

Approximating the Brillouin zone as a cylinder of the volume $2\pi
^{2}q_{d}^{2}/c$ with the Debye momentum $q_{D}=2\sqrt{\pi }/a$ and
integrating over azimuthal (in-plane) angle yield
\begin{equation}
{\frac{V(r)}{{4E_{c}}}}=\int_{0}^{1}\int_{0}^{1}{\frac{dtdyJ_{0}(\sqrt{2\pi}
rt)t^{5}(y^{2}+\eta t^{2})}{{[t^{2}(y^{2}+\eta t^{2})+\zeta (t^{2}
-\Re)]^{2}+\zeta ^{2}\Im^{2}}}},  \label{final}
\end{equation}%
where $\Re=R(kt,t^{2}/2-\tilde{u},k^{2}\tilde{\beta})+R(kt,t^{2}/2+\tilde{u}%
,k^{2}\tilde{\beta})$, $\Im=I(kt,t^{2}/2-\tilde{u},k^{2}\tilde{\beta}%
)-I(kt,t^{2}/2+\tilde{u},k^{2}\tilde{\beta})$, $E_{c}=e^{2}c/(2\pi
^{2}\epsilon _{0}\kappa a^{2})\approx 0.71\mathrm{eV}$ (with $\epsilon
_{\infty }=5$ and $\epsilon _{s}=30$ of La$_2$CuO$_4$ \cite{alebra}). The
on-site, the nearest-neighbor and the next-nearest-neighbor attractions
correspond to $r=0,1,\sqrt{2}$, respectively. $J_{0}(x)$ is the Bessel
function, $k=k_{F}/q_{D}$ is the dimensionless Fermi momentum, $\tilde{u}%
=\omega _{0}m^{\ast }a^{2}/4\pi \hbar \approx 0.015(m^{\ast }/m_{e})$ (for $%
\hbar \omega _{0}=80$ meV), $\zeta =e^{2}cm^{\ast }/(\epsilon _{0}\kappa \pi
^{3}\hbar ^{2})\approx 1.23(m^{\ast }/m_{e})$, and $\eta =4c^{2}/\pi
a^{2}\approx 3.93$. If one assumes that carriers are scattered off
impurities with the density equal to the carrier density, as in La$_{2-x}$Sr$%
_{x}$CuO$_{4}$, then in the Born approximation $\tilde{\beta}=\beta
_{0}(m^{\ast }/m_{e})^{2}$ with $\beta _{0}$ independent of the carrier mass
and density. Using $k_{F}l\approx 20$ and $m^{\ast }=2m_{e}$ as found in YBa$%
_{2}$Cu$_{3}$O$_{6.5}$\cite{und} yields $\beta _{0}\approx 0.0125$.

We can now evaluate the doping ($x$) dependence of $V(r)$ and $m^{\ast }$ by
just changing the dimensionless Fermi momentum, $k=(x/2)^{1/2}$, in Eq.(\ref%
{final}) and solving Eqs (\ref{final},\ref{mass},\ref{g}) selfconsistently
with any bare band mass (we choose $m=m_{e}$ in our numerical calculations).
At very low doping the on-site and nearest-neighbor inter-site attractions
are enormous, $V(0)\approx 1.25$ eV and $V(1)\approx 0.87$ eV, respectively,
and carriers are rather heavy, $m^{\ast }/m\approx 10$. Such heavy polarons
are readily localised by disorder accounting for the Mott variable range
hopping, which explains conduction of lightly doped cuprates.

With doping, the attraction and the polaron mass drop, Figs.(\ref%
{attractionfig},\ref{massfig}), respectively. However, the on-site ($r=0$)
and the inter-site ($r=1$) attractions are well above the superexchange
(magnetic) interaction $J$ (about 100 meV) in underdoped and optimally doped
compounds since the non-adiabatic carriers cannot perfectly
screen high-frequency electric fields. Both attractions and the mass renormalization remain also
substantial at overdoping.  The polaron mass, Fig.(\ref{massfig})
agrees reasonably well with the experimental masses \cite{und,over}.
Increasing the phonon frequency enhances the attraction and lowers the
polaron mass in underdoped compounds with little effect on both quantities
at overdoping, Figs.(\ref{attractionfig},\ref{massfig}). Decreasing
(increasing) the band mass makes polarons lighter (heavier).

In conclusion, we have quantified the carrier-carrier attraction and mass
renormalisation induced by EPI in layered doped ionic lattices. The Fr\"{o}%
hlich EPI with high-frequency optical phonons turns out to be the key
pairing interaction in underdoped cuprates and remains essential at
overdoping. What is more surprising is that EPI is clearly beyond the BCS-ME
approximation since its magnitude is larger or comparable with the Fermi
energy and the carriers are in the non-adiabatic or near-adiabatic regimes
depending on doping. Both conditions point to a crossover from the
bipolaronic to polaronic superconductivity \cite{ale83} with doping.

We thank Janez Bon\'{c}a, Jozef Devreese, Viktor Kabanov, Dragan Mihailovi%
\'{c}, and Stuart Trugman for stimulating discussions. This work was
partially supported by the Royal Society.


\begin{thebibliography}{99}
\bibitem{ref00} $\omega _{p}$ represents the plasma frequency at high values
of the carrier density, when $\omega _{p}>\omega _{0}$ . In the other limit
of low density, the ions screen the carrier plasma oscillations, so that the
plasma frequency contains $\epsilon _{s}$ rather than $\epsilon _{\infty }$.
\end{thebibliography}
\end{document}